# Topologically protected strong coupling and entanglement between distant quantum emitters

**Yujing Wang[1], Jun Ren[2], Weixuan Zhang[1], Lu He[1], and Xiangdong Zhang[1*]**

[1]Key Laboratory of advanced optoelectronic quantum architecture and measurements of Ministry of Education, School of Physics, Beijing Institute of Technology, 100081, Beijing, China

[2]College of Physics and Hebei Key Laboratory of Photophysics Research and Application, Hebei Normal University, 050024, Shijiazhuang, Hebei, China

[*]Corresponding author: zhangxd@bit.edu.cn

The realization of robust strong coupling and entanglement between distant quantum emitters (QEs) is very important for scalable quantum information processes. However, it is hard to achieve it based on conventional systems. Here, we propose theoretically and demonstrate numerically a scheme to realize such strong coupling and entanglement. Our scheme is based on the photonic crystal platform with topologically protected edge state and zero-dimensional topological corner cavities. When the QEs are put into topological cavities, the strong coupling between them can be fulfilled with the assistance of the topologically protected interface state. Such a strong coupling can maintain a very long distance and be robust against various defects. Especially, we numerically prove that the topologically protected entanglement between two QEs can also be realized. Moreover, the duration of quantum beats for such entanglement can reach several orders longer than that for the entanglement in a conventional photonic cavity, making it be very beneficial for a scalable quantum information process.

# I. INTRODUCTION

The controllable interaction between quantum emitters (QEs) is an essential ingredient for the realization of scalable quantum information systems [1, 2]. For example, the realization of large-scale quantum networks requires the capability to interconnect many 'quantum nodes', where each of them should consist of a micro-resonator and a set of trapped atoms [3, 4]. It is necessary to control the coupling of atoms with micro-resonators and the interaction between atoms in various micro-resonators. At the same time, it also requires scientific capabilities for generating quantum coherence and entanglement between distant atoms. Thus, in the past few years, many investigations have been done to construct strong coupling and interactions between QEs [5-14]. It is found that the strong coupling and entanglement between two QEs can be easily realized when they are placed very close to one another [15-20]. However, it is very difficult to be achieved when they are far away from each other (for example, tens or hundreds of wavelengths) [21-25]. Besides, even if the strong coupling and entanglement between distant QEs can be realized, how to keep them stable and free from environmental perturbations is also a problem. That is, how to realize the robust strong coupling and entanglement between distant QEs is an open problem.

Recent developments in topological photonics [26-29], especially in topological quantum optics [30-40], have made it possible to solve the above problem. By introducing the topology into optics, some attractive properties, such as the backscattering-immune propagation of photonic edge modes [41-50], can be realized. The combination of topology and quantum mechanics can bring more interesting phenomena [30-40], including the topological quantum optics interface [30], topological sources of quantum light [31-33], topologically robust transport of entangled photons [34], quantum interference of topological states of light [35] and so on. Very recently, a new class of higher-order topological insulators have been proposed and experimentally demonstrated in many different systems [51-70, 76]. The zero-dimensional (0D) topological corner state has been observed in the two-dimensional (2D) photonic crystal (PhC) slab [57, 58, 69, 70]. Such a high-order topological corner state provides an ideal platform to design topological nanocavity [69-70].

Motivated by the above investigations, in this work we provide a scheme to realize the robust strong coupling and entanglement between distant QEs in the PhC platform. The designed PhC platform possesses topologically protected edge states and 0D topological corner cavities. When the QEs are put in topological cavities, they are coupled strongly and entangled through the topologically protected edge state. Such strong coupling and entanglement can always exist even though the QEs are separated by a long distance, and they are robust against disorders.

## II. MODELS AND SYSTEMS

A simplified top view of our scheme is shown in Fig. 1(a), where the designed PhC platform contains five kinds of integrable modules (I-V) in different regions. The two crossing points are two symmetrically distributed topological corner cavities $C_1$ and $C_2$ with identical parameters. A topologically protected interface state as shown by the solid red line is designed near two cavities. The PhC sample is assembled inside a waveguide with two pair of high-reflectivity metallic plates ($y$ and $z$ directions) by golden color planes, shown in Fig. 1(b). The metal plates in the $y$-direction ($M_1$ and $M_2$) as marked by orange lines are put on the left and right boundaries to ensure the density of interface states of the waveguide modes modified (or discretized) to form Fabry-Perot (FP) waveguide modes. In this case, the corner state and interface state coexist, and the indirect coupling between distant quantum emitters through the forced oscillations of the Fabry-Perot (FP) waveguide modes can be achieved. Here, we consider the transverse magnetic mode with the out-of-plane electric field and the in-plane magnetic field. The distance between $M_1$ ($M_2$) and $C_1$ ($C_2$) is marked by $L_1$ ($L_3$), and the distance between two cavities is expressed as $L_2$. Note that the role of the metal plates $M_1$ and $M_2$ (equivalent to two mirrors) is to ensure that the density of states of the waveguide modes is modified (or discretized) to form appropriate Fabry–Perot (FP) waveguide modes, which can help realize the indirect strong coupling between indirect coupling and entanglement between the distant quantum emitters. Indeed, it may excite the unwanted FP resonance if we can't design the length of the waveguide $L_1 + L_2 + L_3$ properly, which would make the strong coupling and entanglement impossible. Thus, the mode-mismatching conditions need to be satisfied, which has been introduced in Ref. [77]. It is worthy to note that such a design is universal, meaning that similar phenomena can be observed in different wavelength regions (from near-infrared to microwave) by suitably scaling the structural parameters. Without loss of generality, we study the phenomena in the near-infrared region, and $Al_2O_3$ ($\varepsilon=7.5$) cylinders with various diameters are used to form the whole system, where all of the five regions possess triangular lattices with the lattice constant $a$ being 1.25 μm and the unit cell contains six $Al_2O_3$ cylinders. The heights of these cylinders $h$ are taken as $0.1a$.

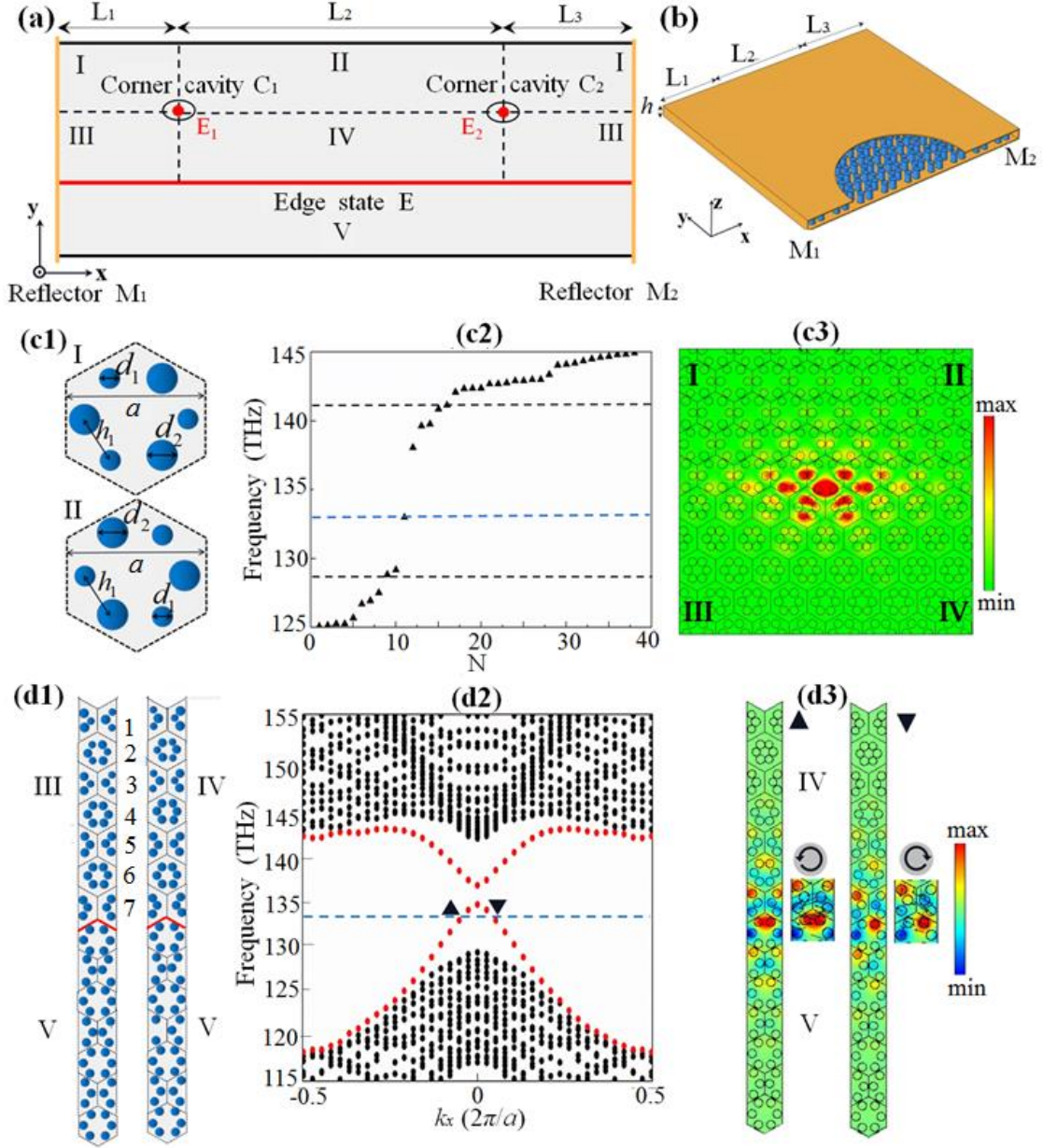

FIG. 1. (a) Simplified top view of the PhC system including two distant quantum emitters $E_1$ and $E_2$ with electric moments $\vec{\mu}_1$ and $\vec{\mu}_2$ (parallel to cylinders). Two black circles represent topological corner cavities, while the red line indicates a topological interface state. The orange lines represent two high-reflectivity metal plates (M1 and M2). (b) Schematic view of the designed PhC system. Two pairs of parallel metal plates are shown by golden color planes. (c) Illustration of the topological corner state. The (c1) is the lattices in regions I and II. The diameter of the small (large) cylinder is marked by d1 (c2) and the intra-unit-cell separation

**between neighboring cylinders is h1. The (c2) and (c3) exhibit the eigen-energy spectrum and electric field distribution of the topological corner state, respectively. (d) Calculated band diagram (d2) based on a supercell composed of region IV and V shown in the (d1). The red dotted line represents the topological interface state. (d3) The corresponding distributions of the out-of-plane electric field.**

As schematically shown in Figs. 1(c1) and 1(d1), both the lattices in regions I/II and III/IV satisfy the inversion symmetry. Here the diameter of small (large) cylinder $d_1$ ($d_2$) is set as 0.28 μm (0.32 μm) and the intra-unit-cell separation between neighboring cylinders is $h_1$=0.371$a$. The 0D corner states we designed appear when the sublattice symmetry breaking and lattice deformation coexist, while the design of 1D interface states only needs lattice deformation. In order to integrate the corner cavities and interface state, the size of cylinders in regions III and IV should be gradually varied from the first row to the fifth row. The gradually varied diameter of the small (large) cylinders $d_1$ ($d_2$) 0.28μm (0.32μm), 0.285μm (0.315μm), 0.29μm (0.31μm), 0.295μm (0.305μm), 0.3μm (0.3μm) are used from the first row to the fifth row, and $h_1$ are set as 0.243$a$, 0.243$a$, 0.262$a$, 0.281$a$ and 0.3$a$, respectively. Besides, the parameters of the sixth and seventh rows are the same as those of the fifth row The diameter and the intra-unit-cell separation between neighboring cylinders in region V are taken as 0.3μm and 0.36$a$, respectively.

In such a case, two topological corner cavities can appear at the crossing of four regions marked by black circles in Fig. 1(a), whose quality factors can reach ~$4\times10^6$ in our design. Note that the distance between the two corner cavities is critical to define the topological state inside the bulk region II and IV, and the minimum length of $L_2$ is 4$a$. Figure 1(c2) displays the eigen-energy spectrum of the two corner cavities, a state at 133 THz inside the gap of the topological kink states appears [68]. The corresponding electric field distribution of the state is given in Fig. 1(c3), indicating the existence of the topological corner state. At the same time, a topologically protected interface state with the working frequency located at 133 THz, which corresponds to the resonance frequency of the corner cavity mode, can be realized. This can be demonstrated by the calculated dispersion relation presented in Fig. 1(d2) and the distribution of the electric field plotted in Fig. 1(d3). As seen in Fig. 1(d2), two topological interface states appear in the bulk band gap as highlighted by black triangles. The distributions of the electromagnetic field obtained by an eigenmode analysis at the two points with opposite momenta [upward and downward triangles in Fig. 1(d2)] are displayed in Fig. 1(d3). Note that the QEs put in the cavities below do not consider the chirality, but it can be regarded as the superposition of left-handed light and right-handed light. When they are coupled to the interface state, the left-handed light and right-handed light can propagate to the left and right, respectively.

## III. TOPOLOGICALLY PROTECTED STRONG COUPLING BETWEEN DISTANT QUANTUM EMITTERS

We put a pair of identical QEs $E_1$ and $E_2$ into two topological corner cavities, respectively, as shown in Fig. 1(a). The two-level QEs are sufficiently far from each other so that the Coulomb interaction between them can be ignored. Under the electric-dipole and rotating wave approximations, the Hamiltonian of the QE system can be expressed as

$$H = \sum_{i=1,2} \hbar(\omega_0 + \delta_i)\sigma_i^\dagger \sigma_i + g_{12}\left(\sigma_1^\dagger \sigma_2 + \sigma_2^\dagger \sigma_1\right), \tag{1}$$

where $\sigma_i^\dagger$ and $\sigma_i$ are the creation and annihilation operators of the $i$th QE ($i$=1,2). $\omega_0$ is the transition frequency for both of QEs. $\delta_i$ is the environment-induced Lamb shift, which makes a small contribution to the dynamics when the coherent coupling between two QEs is large [72]. The second term in the Hamiltonian plays a role of coherent coupling between the $E_1$ and $E_2$, and the coherent coupling coefficient, which arises from the dipole-dipole interaction between the QEs through the near-field, is expressed as:

$$g_{12} = \frac{1}{\pi \varepsilon_0 \hbar} \mathcal{P} \int_0^\infty \frac{\omega^2 \operatorname{Im}\left[\vec{\mu}_1^* \cdot \overleftrightarrow{G}(\vec{r}_1, \vec{r}_2, \omega) \cdot \vec{\mu}_2\right]}{c^2(\omega - \omega_0)} d\omega, \tag{2}$$

where $\overleftrightarrow{G}(\vec{r}_i, \vec{r}_j, \omega)$ is the classical Green tensor of the system, which is the solution of the tensor equation $[\nabla \times \nabla \times -k_0^2 \varepsilon(\vec{r}, \omega)]\overleftrightarrow{G}(\vec{r}, \vec{r}'; \omega) = \overleftrightarrow{I}\delta(\vec{r} - \vec{r}')$. Here, $\varepsilon(\vec{r}, \omega)$ is the relative permittivity, and the magnetic response has been omitted. The symbol $\mathcal{P}$ stands for the principle integral, and $\vec{\mu}_1$ and $\vec{\mu}_2$ are dipole moments of two QEs. In our calculations, we have taken $|\vec{\mu}_1| = |\vec{\mu}_2| = |e| \times |r_0|$, where $|r_0|$=1 Å [73]. According to Ref. [73], the principle integral can be avoided by using the Kramers-Kronig relation with the help of the contour integral. In such a case, the coherent term can be simplified as

$$g_{12} = \frac{\omega_0^2}{\varepsilon_0 \hbar c^2} \text{Re}\left[ \vec{\mu}_1^* \cdot \ddot{\vec{G}}^s(\vec{r}_1, \vec{r}_2, \omega) \cdot \vec{\mu}_2 \right]. \tag{3}$$

It should be pointed out that, the designed structures are dielectric materials, and the magnetic effects are very weak, therefore we only consider the electric effects in this work. The Green tensor $\ddot{\vec{G}}^s(\vec{r}_1, \vec{r}_2, \omega)$ showed in Eq. (3) represents the scattering Green tensor of a source dipole located at $\vec{r}_2$ and the field dipole located at $\vec{r}_1$, which can be obtained by the Ref. [77]:

$$\vec{n}_1 \cdot \ddot{\vec{G}}^s(\vec{r}_1, \vec{r}_2, \omega) \cdot \vec{n}_2 = -\vec{n}_1 \cdot \vec{E}_s(\vec{r}_1)|_{\vec{r}_2}, \tag{4}$$

where $\vec{n}_1$ and $\vec{n}_2$ are the direction of the field and source dipole moment respectively, and $\vec{E}_s(\vec{r}_1)|_{\vec{r}_2}$ is the scattering electric field at the location of field dipole $\vec{r}_1$ induced by the source dipole located at $\vec{r}_2$. In this work, the orientation of the source dipole moment is taken as the positive direction of the z-axis, which will excite the TM modes of the designed structure, and the electric direction of the field dipole felt is also parallel to the z-axis. In this case, the Eq. (4) turns to

$$G_{zz}^s(\vec{r}_1, \vec{r}_2, \omega) = -\vec{E}^s(\vec{r}_1)|_{\vec{r}_2}, \tag{5}$$

and Eq. (3) in this work turns to

$$\begin{aligned} g_{12} &= \frac{\omega_0^2}{\varepsilon_0 \hbar c^2} \text{Re}\left[ G_{zz}^s(\vec{r}_1, \vec{r}_2, \omega) |\vec{\mu}_1||\vec{\mu}_2| \right] \\ &= \frac{\omega_0^2}{\varepsilon_0 \hbar c^2} \text{Re}\left[ -E_z^s(\vec{r}_1)|_{\vec{r}_2} \cdot |\vec{\mu}_1||\vec{\mu}_2| \right] \end{aligned}. \tag{6}$$

The component of the scattering electric field along the z-direction $E_z^s(\vec{r}_1)|_{\vec{r}_2}$ can be easily obtained numerically by the finite element method. Besides the coherent coupling, the incoherent interference $\gamma_{12}$ between the two QEs mediated by the environment can be expressed as [73]

$$\gamma_{12} = \frac{2\omega_0^2}{\varepsilon_0 c^2 \hbar} \text{Im}\left[ \vec{\mu}_1^* \cdot \ddot{\vec{G}}(\vec{r}_1, \vec{r}_2, \omega) \cdot \vec{\mu}_2 \right]. \tag{7}$$

The parameter $\gamma_{12}$ showed in Eq. (7) introduces an incoherent coupling between two QEs through the vacuum field, which can affect the spontaneous emission of two QEs. The Green tensor $\ddot{G}(\vec{r}_1, \vec{r}_2, \omega)$ showed in Eq. (7) is the total Green tensor of a source dipole located at $\vec{r}_2$ and the field dipole located at $\vec{r}_1$, which can be written as

$$\ddot{G}(\vec{r}_1, \vec{r}_2, \omega) = \ddot{G}^s(\vec{r}_1, \vec{r}_2, \omega) + \ddot{G}^{vac}(\vec{r}_1, \vec{r}_2, \omega), \tag{8}$$

where $\ddot{G}^{vac}(\vec{r}_1, \vec{r}_2, \omega)$ is the vacuum Green tensor, which is readily known in Ref. [77]. The Green tensor showed in Eq. (8) can be obtained with

$$\vec{n}_1 \cdot \ddot{G}(\vec{r}_1, \vec{r}_2, \omega) \cdot \vec{n}_2 = -\vec{n}_1 \cdot \vec{E}(\vec{r}_1)|_{\vec{r}_2}, \tag{9}$$

where $\vec{E}(\vec{r}_1)|_{\vec{r}_2}$ is the total electric field at the location of field dipole $\vec{r}_1$ induced by the source dipole located at $\vec{r}_2$ and can also be obtained by the finite element method. Thus, the parameter $\gamma_{12}$ showed in Eq. (7) can be written as

$$\begin{aligned}\gamma_{12} &= \frac{2\omega_0^2}{\varepsilon_0 c^2 \hbar} \text{Im}\left[G_{zz}(\vec{r}_1, \vec{r}_2, \omega)|\vec{\mu}_1||\vec{\mu}_2|\right] \\ &= \frac{2\omega_0^2}{\varepsilon_0 c^2 \hbar} \text{Im}\left[-E_z(\vec{r}_1)|_{\vec{r}_2} \cdot |\vec{\mu}_1||\vec{\mu}_2|\right]\end{aligned}, \tag{10}$$

where $E_z(\vec{r}_1)|_{\vec{r}_2}$ is the component of the total electric field along the $z$ direction at the location of $\vec{r}_1$ induced by a source dipole located at $\vec{r}_2$. Since the two QEs locate at the same environment, they possess the identical emission rate, which can be written as

$$\begin{aligned}\gamma_{11} &= \frac{2\omega_0^2}{\varepsilon_0 c^2 \hbar} \text{Im}\left[\vec{\mu}_1^* \cdot \ddot{G}(\vec{r}_1, \vec{r}_1, \omega) \cdot \vec{\mu}_1\right] \\ &= \frac{2\omega_0^2}{\varepsilon_0 c^2 \hbar} \text{Im}\left[G_{zz}(\vec{r}_1, \vec{r}_1, \omega)|\vec{\mu}_1||\vec{\mu}_1|\right]. \\ &= \frac{2\omega_0^2}{\varepsilon_0 c^2 \hbar} \text{Im}\left[-E_z(\vec{r}_1)|_{\vec{r}_1} \cdot |\vec{\mu}_1||\vec{\mu}_1|\right]\end{aligned} \tag{11}$$

where $E_z(\vec{r}_1)|_{\vec{r}_1}$ is the component of the total electric field along the $z$-direction at the location of $\vec{r}_1$ induced by a source dipole located at the same position.

In fact, three physical quantities $g_{12}$, $\gamma_{12}$ and $\gamma_{11}$ are sufficient to characterize the coupling properties between the two QEs. In Fig. 2, we provide the calculated results of these physical quantities for two PhC systems with different distances between two QEs based on the finite element method. Figures 2(a) and 2(b) correspond to the case with $L_2$=18$a$ (10 times of the transition wavelength), and Figs. 2(c) and 2(d) to the case with $L_2$=58$a$ (32 times of the transition wavelength). Figures 2(a) and 2(c) show the decay rate of QEs as a function of the transition frequency for two cases, respectively. It is clearly shown that the splitting of the resonance peaks appears around 133 THz for both cases. The Q factors of these peaks are over $1.0\times10^6$, which indicates the existence of the strong coupling between a single QE and a corner cavity.

It is well known that the spontaneous decay rate $\gamma_{11}$ is proportional to the density of the state. Thus, its split corresponds to the Rabbi split, which is a signal of the strong coupling between two QEs. The red lines in Figs. 2(b) and 2(d) display the calculated results of the normalized coherent coupling coefficient ($g_{12}/\gamma_{11}$) as a function of the transition frequency. The corresponding incoherent coupling coefficient ($\gamma_{12}/\gamma_{11}$) are shown by blue lines. It is seen that the maximum value of $g_{12}/\gamma_{11}$ is 7.184 at $f$=133.0175 THz and -6.003 at $f$=133.0173 THz when $L_2$=18$a$ and 58$a$, respectively. In such cases, $\gamma_{12}/\gamma_{11}$ are near zero, indicating a perfect coherent coupling. It is generally believed that strong coupling can be achieved when $g_{12}/\gamma_{11}$ reaches 5 [74, 75]. Consequently, the present cases show very strong coupling behavior. Comparing these two cases, it is interesting to find that the value of $g_{12}/\gamma_{11}$ decays slowly with increasing $L_2$, which means that such strong coupling can maintain a very long distance, which is much longer than other systems [73].

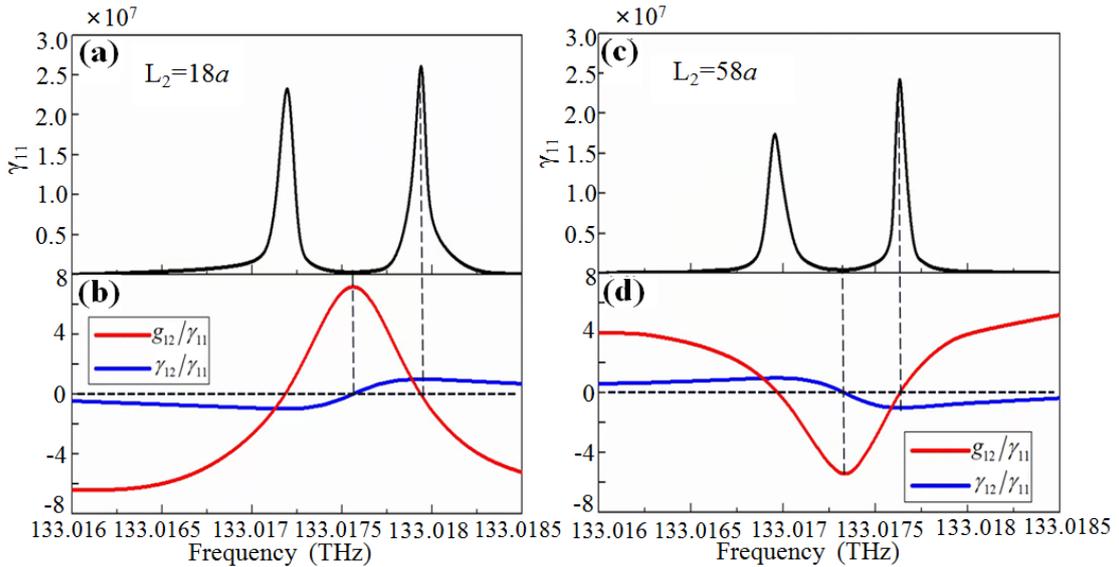

**FIG. 2. Spontaneous decay rates of single QE $\gamma_{11}$ (black lines) for two cases (a) $L_1$=$L_3$=20.5$a$, $L_2$=18$a$ and (c) $L_1$=$L_3$=15$a$, $L_2$=58$a$. (b) and (d) represent the coherent (red lines) and incoherent (blue lines) coupling parameters for these two cases.**

Another important feature is that the above strong couplings are immune to various defects. In order to check the robustness of these strong couplings, we consider two types of defects. One is to introduce intentionally deformed artificial atoms at the interface, the other is to use a sharply bent interface. For the first type of defect, we make the diameters and positions of some cylinders slightly change at the interface between regions III (IV) and V as has been done in Refs. [48-49]. As shown in Fig. 3(c), black cylinders represent their position moved to the left by $0.02a$ and white cylinders indicate their diameters become 0.29 μm. In such a case, the calculated results for the decay rate $\gamma_{11}$ as a function of the transition frequency for $L_2=18a$ are shown in Fig. 3(a). It is clearly shown that the splitting phenomenon of the resonance peak and the Q factor of each peak nearly remain the same. As for the coherent and incoherent coupling coefficients, the maximum value of $g_{12}/\gamma_{11}$ decreases to 6.952 at $f$=133.0175 THz, and the corresponding value of $\gamma_{12}/\gamma_{11}$ is 0.002. In such a case, the strong coherent coupling still exists. In order to analyze the origin of these phenomena, the distribution of the electric field at the interface with disorders is plotted in Fig. 3(c) when a chiral light source (marked with a star) [71] is put on the interface. We find that there is almost no backscattering with disorders. It is the topologically protected channel that leads to robust strong coupling between QEs.

For the second type of defect, we consider a triangle-like interface as shown in Fig. 3(f). Due to the topologically protected characteristics, the unidirectional propagation of the excited electric field appears under such bending, as shown in Fig. 3(f). This makes couplings [see Fig. 3(e)] and splitting of the decay rate [see Fig. 3(d)] less affected by the bent interface. The calculated results are shown in Figs. 3(d) and 3(e). The Rabbi splitting is observed again, and the maximum coherent coupling and the corresponding incoherent coupling coefficients are 6.349 and 0.013. Such little difference of the coefficients compared with the case of the straight interface means that the strong coherent coupling still exists with the bending interface.

In addition, not only the interface state, which is used to assist two quantum emitters to realize the strong coupling, is topologically protected, but also corner cavities are robust against bulk disorders, making the strong coupling between two QEs be also robust against bulk disorders. Thus, the extensive studies on the effect of the random dislocation or different diameter of surrounding cylinders of the corner cavities and the coupling waveguide have been done. We randomly selected a number of cylinders around each element for our study. Here, the ratio of the diameter of the cylinder after the change to that before the change is expressed as $D$. Take the case of $L_2 = 18a$ as an example. When the ratio $D$ is chosen from [0.95, 1.06] in a random way, and positions of some cylinders are shifted randomly by distances chosen from [0.01$a$, 0.05$a$], and the average maximum coherent coupling coefficient can sill reach 6.539. This result is smaller than that without considering the random dislocation or different diameter of cylinders, but the phenomenon of the Rabbi splitting and strong coupling still exist, which exhibit good robust properties.

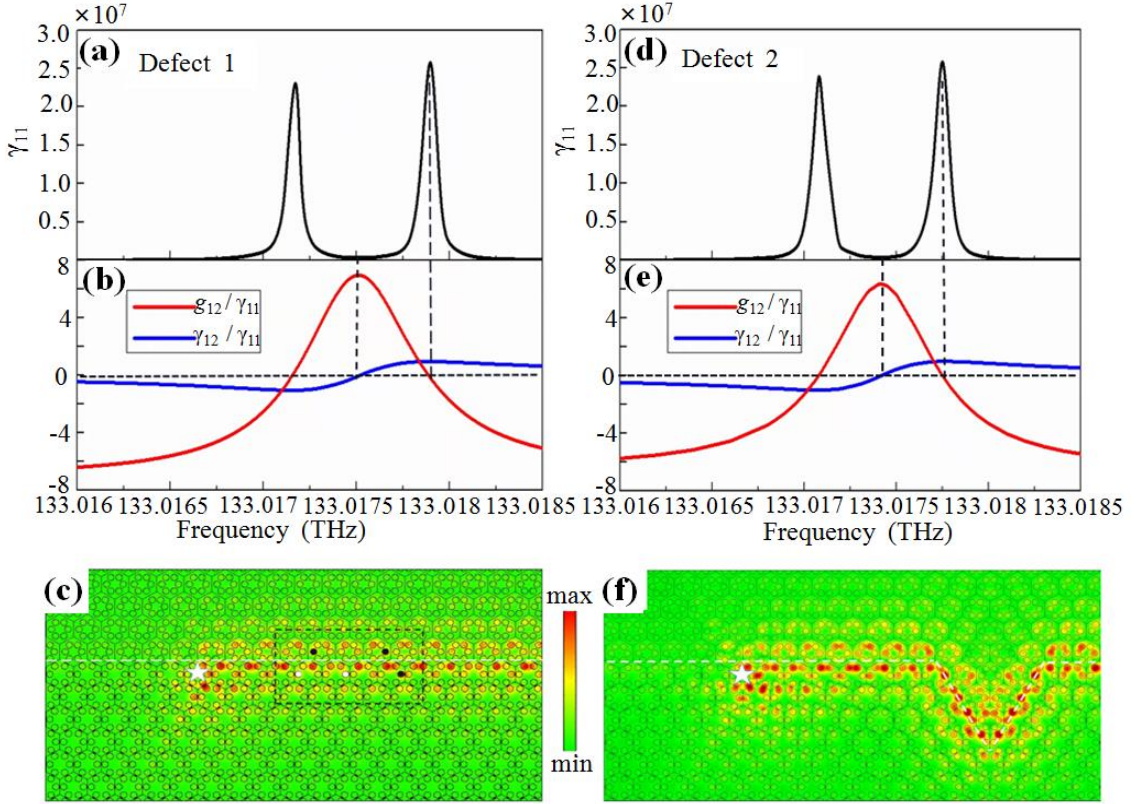

FIG. 3. Spontaneous decay rates of single QE $\gamma_{11}$ (black lines) for two defects, (a) defect 1 and (d) defect 2; (b) and (e) represent the coherent (red lines) and incoherent (blue lines) parameters for the cases of two defects; (c) and (f) are simulated unidirectional propagation of the state along the interface for the two defects respectively. The stars represent chiral light sources near the interface.

## IV. TOPOLOGICALLY PROTECTED ENTANGLEMENT BETWEEN DISTANT QUANTUM EMITTERS

Based on our designed PhC platform, the long-range entanglement between two QEs can also be realized. Initially, we excite one of the QEs. Due to the strong coherent coupling between the two QEs, the excitation transfers back and forth between the two QEs, and entanglement will occur in this process. The dynamic evolution of two QEs can be described by the master equation [21] and we use concurrence to measure two-qubit entanglement between them in this work. Based on the calculated coupling parameters, we can analyze the entanglement dynamics between two QEs by calculating the population and concurrence of the system. The detailed process and definitions of the population and concurrence are given below.

The dynamic evolution of two QEs can be described by the following master equation [21]

$$\partial_t \rho = \frac{i}{\hbar}[\rho, H] + \sum_{i=1,2} \frac{\gamma_{11}}{2}\left(2\sigma_i \rho \sigma_i^\dagger - \sigma_i^\dagger \sigma_i \rho - \rho \sigma_i^\dagger \sigma_i\right)$$
$$+ \frac{\gamma_{12}}{2}\left(2\sigma_1 \rho \sigma_2^\dagger - \sigma_1^\dagger \sigma_2 \rho - \rho \sigma_1^\dagger \sigma_2\right) \qquad , \qquad (12)$$
$$+ \frac{\gamma_{21}}{2}\left(2\sigma_2 \rho \sigma_1^\dagger - \sigma_2^\dagger \sigma_1 \rho - \rho \sigma_2^\dagger \sigma_1\right)$$

where $\rho$ is the density matrix of two QEs and it is a $4\times 4$ matrix under the normal basis ($\{|00\rangle, |01\rangle, |10\rangle, |11\rangle\}$). The matrix element $\rho_{22(33)}$ represents the population when the 1st (2nd) QE is in the excited state and the 2nd (1st) QE is in the ground state. Here, we label these two matrix elements as $P_1$ and $P_2$, respectively. It should be pointed out that the Lindblad equation requires the Markov approximation. In this work, the frequencies we choose are that the coherent coupling between the two QEs is strong, but the coherent coupling between the single QE and the field is weak, which justifies the reasonable use of the Markov approximation. We use two-qubit concurrence to measure the entanglement between two QEs, which is defined as [78]

$$C(\rho) = \max\{0, \ \lambda_1 - \lambda_2 - \lambda_3 - \lambda_4\}, \qquad (13)$$

where $\lambda_i$ s are the eigenvalues of the matrix $\sqrt{\sqrt{\rho}\tilde{\rho}\sqrt{\rho}}$ in decreasing order and $\tilde{\rho}$ is the spin-flipped state, which is defined as

$$\tilde{\rho} = (\sigma_y \otimes \sigma_y)\rho^*(\sigma_y \otimes \sigma_y), \qquad (14)$$

and $\rho^*$ is the complex conjugate of the matrix $\rho$.

In Fig. 4, we provide the calculated results of the population and concurrence as a function of the time for the case with $L_2$=18$a$ and 58$a$. Here the frequency is taken as 133.0175 THz and 133.0173 THz, respectively, which correspond to the peaks of the coherent coupling $g_{12}/\gamma_{11}$ shown in Figs. 2(b) and (d). Figures 4(a) and (b) correspond to the calculated results of the population for the two cases, respectively, and Figs. 4(c) and (d) represent the corresponding time-resolved entanglement concurrence. Initially, the first QE is excited and the second is in its ground state, which correspond to the populations $P_1$=1 and $P_2$=0. Here $P_1$ and $P_2$ are the diagonal elements of the density matrix of two QEs. After the $E_1$ is excited, the population of $E_2$ increases at the beginning and oscillates due to the large coherent coupling, and the excitation transfers between the two QEs back and forth. A clear Rabbi oscillation is found, and the high extinction ratio indicates clearly the successful exchange of photons between the distant QEs. The Rabbi oscillation is seen to continue for more than 12 μs, demonstrating the long coherence time of photons in this system. Comparing the two cases, it is found that both the coherent time and the attenuation amplitude of oscillation have little change with the increase of $L_2$, which means that such a Rabbi oscillation can maintain a very long distance.

According to Figs. 4(b) and 4(d), the quantum beat frequency is nearly equal in both cases. For example, when the time $t$=4 μs, the concurrence for two cases reach the 6th beat and the peak value of the concurrence only decreases from 0.3 to 0.25 when the distance between two cavities increases from 18$a$ to 58$a$. And the entanglement duration remains almost unchanged. This proves that the increase in distance does not have much effect on the persistence of entanglement. Comparing the present results with the concurrence of two-qubit system in a photonic waveguide [21], we find that the duration of quantum beats for such a case can reach several orders longer than that for the previous cases.

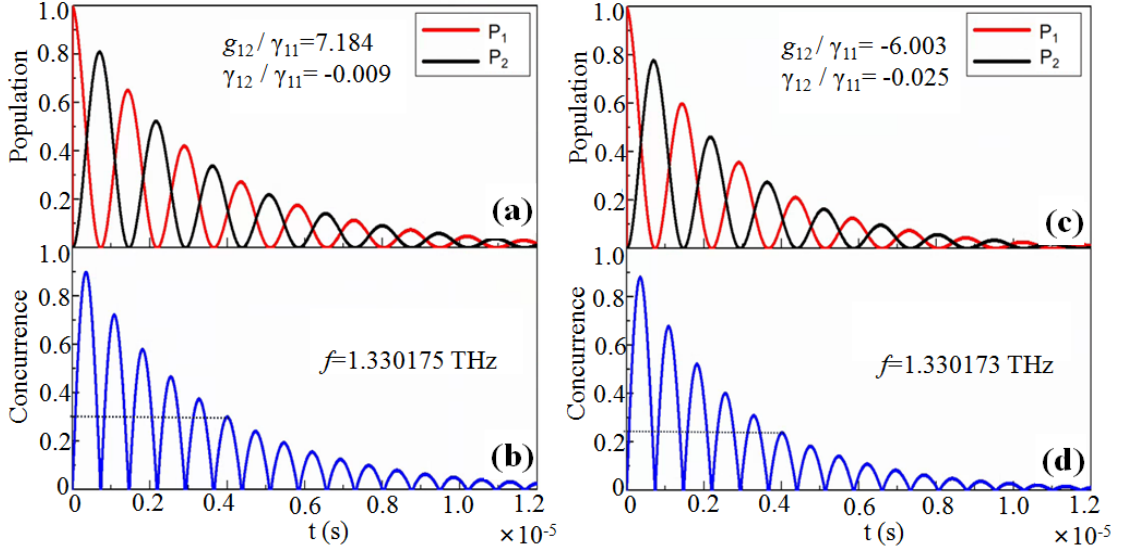

**FIG. 4. (a)** Populations and entanglements of two QEs when they are inserted in the corner of the designed system [see Fig. 1(a)]. Here $L_1=L_3=20.5a$ and $L_2=18a$. **(b)** The corresponding populations and entanglements for the case of $L_1=L_3=15a$ and $L_2=58a$.

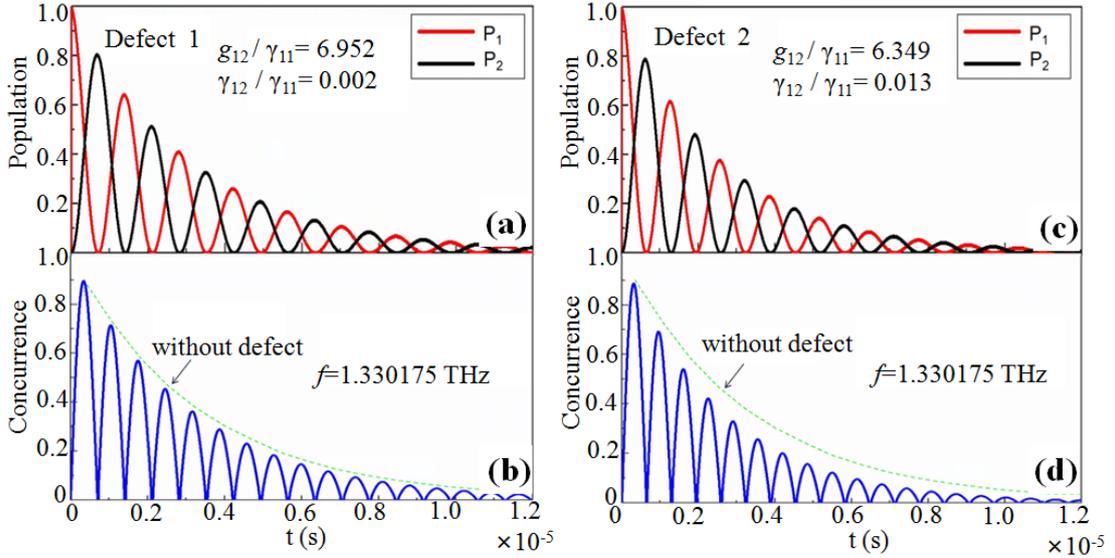

**FIG. 5.** Populations (a) and entanglement concurrence (b) of two QEs as a function of time for the case of the first defect with $f= 133.0175$ THz, $g_{12}/\gamma_{11} =6.952$ and $\gamma_{12}/\gamma_{11} =0.002$. The curves in (c) and (d) have the same meaning with (a) and (b) for the case of the second defect with $f= 133.0175$ THz, $g_{12}/\gamma_{11} =6.349$ and $\gamma_{12}/\gamma_{11} =0.013$.

Another advantage of constructing entanglement in this way is that it is topologically protected, that is, it is robust against the disorder. Figures 5(a) and 5(b) show the calculated results of the population and concurrence for the first type of disorder (deformed artificial atoms at the interface), while the corresponding results for the second type of disorder (sharply bent interface) are plotted in Figs. 5(c) and 5(d). Here $L_2=18a$ is taken for two types of disorder. For comparison, the peak envelopes of concurrence without defects are outlined by green dotted lines in Fig. 5(b) and 5(d). It can be seen clearly that the basic characteristics of entanglement dynamics remain unchanged, but the peak values of concurrence decrease a little under the influence of two kinds of defects. This means that robust entanglement with very long distances can be constructed, which is very important to quantum information processing.

## V. DISCUSSION AND CONCLUSION

The above results are obtained without considering the properties of realistic quantum emitters but only considering their idealized choice, but this does not mean that our scheme is difficult to achieve. In fact, the experiment of the coupling between the second-order topological corner state and single quantum emitters has been realized [69]. In addition, the orientation of the dipole moments is taken as the positive direction of the z-axis, where topologically protected strong coupling and entanglement between distant quantum emitters have been achieved. In order to make our results more complete, we further study the influence of the direction of dipole moments not along the z-axis on the results. The results showed that the effect was not obvious. For example, the maximum coherent coupling coefficient can still reach over 7 when the direction of dipole moments deviates from the z-axis by 10° in the case of $L_2=18a$.

In conclusion, the strong coupling between two QEs at a very long distance has been realized using the designed a PhC platform with the topologically protected edge state and 0D corner cavities. Besides, we numerically prove that such a strong coupling is topologically protected and robust against the disorder. Furthermore, we have demonstrated that the topologically protected entanglement between two QEs can be also realized. The duration of quantum beats for such entanglement can reach several orders longer than that for the entanglement in a conventional photonic cavity. These results are very important for the scalable quantum information process, such as the quantum network and teleportation, quantum cryptographic, quantum dense coding and parallel computing. Of course, providing a full protocol or a discussion of initialization, control of the interaction, and read out /demonstration is also interesting and important work, and this is also the further research we are going to carry out.

## ACKNOWLEDGMENTS

This work was supported by the National key R & D Program of China (2017YFA0303800) and the National Natural Science Foundation of China through Grant Nos. 91850205 and 11904078. J.R was also supported by Hebei NSF (A2019205266) and Hebei Normal University (L2018B02).